\documentclass[aps,prb,10pt,twocolumn,superscriptaddress]{revtex4-1}
\usepackage{amsmath}
\usepackage{amsfonts}
\usepackage{amssymb}
\usepackage{graphicx}
\usepackage{color}
\usepackage{hyperref}
\hypersetup{colorlinks=true,allcolors=blue}
\usepackage{textcomp}

\begin{document}

\title{From strong to weak temperature dependence of the two-photon entanglement 
resulting from the biexciton cascade inside a cavity}

\author{T. Seidelmann}
\affiliation{Lehrstuhl f{\"u}r Theoretische Physik III, Universit{\"a}t Bayreuth, 95440 Bayreuth, Germany}
\author{F. Ungar}
\affiliation{Lehrstuhl f{\"u}r Theoretische Physik III, Universit{\"a}t Bayreuth, 95440 Bayreuth, Germany}
\author{M. Cygorek}
\affiliation{Department of Physics, University of Ottawa, Ottawa, Ontario, Canada K1N 6N5}
\author{A. Vagov}
\affiliation{Lehrstuhl f{\"u}r Theoretische Physik III, Universit{\"a}t Bayreuth, 95440 Bayreuth, Germany}
\affiliation{ITMO University, St. Petersburg, 197101, Russia}
\author{A. M. Barth}
\affiliation{Lehrstuhl f{\"u}r Theoretische Physik III, Universit{\"a}t Bayreuth, 95440 Bayreuth, Germany}
\author{T. Kuhn}
\affiliation{Institut f{\"u}r Festk{\"o}rpertheorie, Universit{\"a}t M{\"u}nster, 48149 M{\"u}nster, Germany}
\author{V. M. Axt}
\affiliation{Lehrstuhl f{\"u}r Theoretische Physik III, Universit{\"a}t Bayreuth, 95440 Bayreuth, Germany}

\begin{abstract}
We investigate the degree of entanglement quantified by the concurrence of photon pairs that are simultaneously emitted
in the biexciton-exciton cascade from a quantum dot in a cavity.
Four dot-cavity configurations are compared that differ with respect
to the detuning between the cavity modes and the quantum dot transitions, 
corresponding to different relative weights of direct two-photon and sequential single-photon processes.
The dependence of the entanglement on the exciton fine-structure splitting $\delta$
is found to be significantly different for each of the four configurations. 
For a finite splitting and low temperatures, the highest entanglement is found when the cavity modes are in resonance with 
the two-photon transition between the biexciton and the ground state and, in addition, the biexciton has a finite binding energy
of a few meV. However, this widely used configuration is rather strongly affected by phonons such that
other dot-cavity configurations, that are commonly regarded as less suited for obtaining high degrees of entanglement,
become more favorable already at temperatures on the order of $10\,$K and above.
If the cavity modes are kept in resonance with one of the exciton-to-ground-state transitions and the biexciton binding energy is finite,
the entanglement drastically drops for positive $\delta$ with rising temperatures when $T$ is below $\simeq$ 4 K, but is virtually 
independent of the temperature for higher $T$.
\end{abstract}

\maketitle

\section{Introduction}
\label{sec:introduction}

Entangled photon pairs can be used as the fundamental building blocks for a wide
range of applications in quantum communications, quantum cryptography, or quantum
computation\cite{akopian2006, RevModPhys.84.777, RepProgPhys.80.076001}.
Furthermore, entanglement can be used to test fundamentals of quantum mechanics, e.g., by 
revealing violations of Bell's inequality \cite{akopian2006,joens2017}.
Different devices and protocols for the generation of entangled photon pairs have been proposed.
A well-established and especially attractive way of producing (polarization) entangled photon pairs is
the emission of photon pairs via the biexciton cascade in semiconductor quantum dots (QDs) inside a microcavity
which enhances the light collection  efficiency \cite{stevenson06,Young:2006,Hafenbrak:2007,dousse:10,delValle:2011,Mueller:2014,winik2017,Benson2000}. 
One special advantage of using semiconductor quantum dots is the possibility to generate triggered \cite{stevenson06,Young:2006,Hafenbrak:2007} 
or even on-demand \cite{johne2008,Mueller:2014,winik2017} entangled photon pairs
which is of utmost importance for applications.

Entanglement generation from the biexciton cascade is possible 
since the biexciton can decay via two paths, first
into one of the two exciton states and a photon which can be
either polarized horizontally ($H$) or vertically ($V$).
Subsequently, the exciton generated in the first step
can further decay to the QD ground state by emitting
a second photon with the same polarization as the photon
generated in the biexciton decay.
Ideally, the two paths are fully symmetric and the corresponding quantum
state is a coherent superposition of the respective amplitudes,
resulting in a maximally entangled two-photon state.
However, when which-path information is introduced by disturbing the symmetry, e.g.,
by a finite fine-structure splitting between the intermediate exciton states,
the superposition becomes asymmetric and the entanglement decreases.
In principle, it is possible to come close to maximal entanglement in
current experiments, either by selecting QDs which naturally have a
sufficiently small fine-structure splitting \cite{Hafenbrak:2007,Mueller:2014},
by tuning the splitting with external fields \cite{stevenson06,Young:2006,stevenson2006b},
or by applying strain \cite{zhang2015}. 
However, these requirements are rather restrictive.
Therefore, it has been proposed to look for less demanding conditions which still
allow for a high degree of entanglement. 
A prominent proposal of this type considers QDs with a sizable biexciton binding energy which are embedded in a microcavity. 
Besides the possibility of an increased light extraction efficiency, a microcavity offers the advantage that the
resonance between the cavity modes and electronic transitions in the dot can be used to enhance, e.g., direct two-photon transitions
between the biexciton and the ground state compared to sequential transitions from the biexciton to
the exciton or from the exciton to the ground state. 
Since the direct two-photon transitions do not involve the occupation of exciton states,
the fine-structure splitting is effectively not probed, leading
to drastically reduced which-path information and therefore increased entanglement \cite{Schumacher:2012,EdV}.
When the cavity mode is tuned to the two-photon resonance, a finite biexciton binding energy is typically favorable
for entanglement since it shifts the sequential 
single-photon transitions further away from resonance.

In order to systematically compare different configurations
of cavity and QD transition frequencies, a measure for the entanglement is required.
A widely accepted measure is the concurrence, which has a one-to-one correspondence to
the entanglement of formation \cite{Wootters:2001}.
The latter represents the amount of pure-state entanglement 
that is at least present in a mixed state described by a given density
matrix. The concurrence has the advantage that it can be directly
calculated from the values of the reduced density matrix of the
bipartite system for which the entanglement is to be measured \cite{Wootters:1998}.
Here, we focus on the concurrence of simultaneously emitted photon pairs which, albeit
yielding lower signals due to filtering only photons with equal emission times from the cavity,
typically show the highest degree of entanglement in experiments \cite{PhysRevLett.101.170501,Bounouar18}, 
as well as theoretical calculations \cite{EdV,new_paper}.

Phonons are known to have a tremendous impact on the dynamics of
QDs in general \cite{Knorr2003,RabiRevival,ramsay:10,ramsay:10a,PI_phonon-assisted_biexc_prep,Finley_phonon-assisted,Biex-exact,
Jakubczyk2016,BiexPrep,Doris2017,McCutcheonNazir_beyondWC} and on QD-cavity systems in particular 
\cite{Wilson-Rae2002,hohenester09,polariton-phonon,kaer10,Kaer12,Hughes_master_equation_polaron_cavity,PI_Cavity_Soergel,roy12,
HughesCarmichael2013,Iles-Smith2017}. 
Since the pure dephasing induced by acoustic phonons is a major source
of decoherence in QDs \cite{RabiRevival,ramsay:10,ramsay:10a}, 
phonons might also limit the entanglement of the two-photon states generated
in the biexciton cascade. However, in many studies of the entanglement phonons
have either been completely disregarded \cite{johne2008,EdV} or accounted for by
a phenomenological pure dephasing rate \cite{PhysRevB.74.235310,carmele2011,Schumacher:2012}. 
The description based on rates ignores that phonon-induced pure dephasing leads only
to a partial loss of coherence which is nonexponential \cite{dotspektren:02,besombes:01}
and is the origin of many other non-Markovian effects \cite{kaer10,inideco,thorwart2005}.
Furthermore, with phenomenological rates the temperature dependence of the degree of entanglement
cannot be predicted.
An explicit treatment of the phonon impact on the concurrence in the biexciton cascade 
has been presented in Ref.~\onlinecite{harouni14}. However, that paper concentrates on the
contributions from the sequential decay of the biexciton via intermediate excitons
and misses the competition with the direct two-photon decay to the ground state,
which is at the heart of the protocol based on resonant two-photon transitions 
in systems with finite biexciton binding energies proposed in Ref.~\onlinecite{Schumacher:2012}.
The effect of phonons on the concurrence in the case where two-photon transitions dominate the biexciton 
decay has been analyzed in Ref.~\onlinecite{heinze2017} where, however,
no selection of simultaneously emitted photon pairs has been considered. 
As mentioned above, the latter is more favorable for obtaining a high degree of photon entanglement.

In this paper, we investigate the phonon impact on the degree of two-photon polarization entanglement obtained after the decay of a biexciton in a cavity as measured by the concurrence of simultaneously emitted photon pairs. 
We present a comprehensive comparison of representative configurations of cavity and QD transition frequencies referring to physical situations with different relative importance of two-photon and sequential single-photon pathways, respectively.
We find that the phonon influence in combination with the competition between two-photon and one-photon processes leads to strikingly different dependences on the exciton splitting as well as strongly different temperature dependences.

Tuning the cavity to the two-photon resonance and considering a quantum dot with
a biexciton binding energy of a few meV is likely to be the most widely studied
configuration in the literature because it is commonly expected to yield the highest
two-photon entanglement at finite fine-structure splitting.
Indeed at low temperatures we confirm this expectation.
The main result of the present paper is, however, that the distinction of the two-photon resonant configuration
with finite biexciton binding energy to yield the highest concurrence is lost typically already at temperatures
as low as $\sim$ 10 K.

The article is structured as follows. In Sec.~\ref{sec:model} we specify the model and the method used.
We discuss the concurrence of simultaneously emitted photon pairs as the measure of choice when high degrees of entanglement are targeted and explain how this quantity is extracted from the numerical calculations.
In Sec.~\ref{sec:configurations} four configurations with different resonance settings and biexciton binding energies are introduced  which enable us to analyze most clearly the competition between direct two-photon and sequential single-photon processes and its impact on the degree of entanglement.
In Sec.~\ref{sec:results} we demonstrate that the phonon impact strongly depends on the considered configuration, resulting in substantially different dependences on the fine-structure splitting and the temperature.
Deviations from the standard bell shape dependence on the splitting or asymmetries reflect the competition between single- and two-photon processes.
Finally, in the Conclusion, Sec.~\ref{sec:conclusion}, we present a brief summary of the main results of this article.

\section{Theoretical approach}
\label{sec:model}

\subsection{Model}
\label{subsec:model}

We consider a semiconductor QD embedded in a microcavity which is initially
prepared in the biexciton state. 
The dynamics of the statistical operator of the system $\hat{\rho}$ is determined by the Liouville-von
Neumann equation
\begin{equation}
\label{eq:dynamic_density_matrix}
\frac{\mathrm{d}}{\mathrm{d}t}\hat{\rho}=-\frac{i}{\hbar}\left[\hat{H},\hat{\rho}\right]+\mathcal{L}\left[\hat{\rho}\right]
\end{equation}
where $[\,,]$ denotes the commutator. 
The Hamiltonian
\begin{equation}
\label{eq:H}
\hat{H}=\hat{H}_\text{QD-cav}+\hat{H}_\text{QD-phon}
\end{equation} 
takes into account the interaction between the QD and two
linearly polarized cavity modes ($\hat{H}_\text{QD-cav}$) as well as a pure dephasing
type coupling to a continuum of longitudinal acoustic (LA) phonons
($\hat{H}_\text{QD-phon}$). The Lindblad operator
$\mathcal{L}\left[\hat{\rho}\right]$ allows the inclusion of non-Hamiltonian
dynamics, i.e., cavity losses due to for example imperfect mirrors. Thus, the
model contains three parts, which are discussed separately in the following.  

The first part describes the coupling between the QD and two linearly polarized
cavity modes and is modeled by the Hamiltonian \cite{heinze2017}
\begin{equation}
\label{eq:H_QD-cav}
\begin{split}
\hat{H}_\text{QD-cav}=\,&\hbar\omega_H| X_H\rangle\langle X_H|+\hbar\omega_V| X_V\rangle\langle X_V|
\\&+\hbar(2\bar{\omega}_{X}-\omega_B)| B\rangle\langle B|+\sum_{\ell=H,V}\hbar\omega_\ell^c\hat{a}_\ell^\dagger\hat{a}_\ell + \hat{\mathcal{X}},
\end{split}
\end{equation}
where the interaction part is given by
\begin{equation}
\label{eq:QD-cav-coupling}
\begin{split}
\hat{\mathcal{X}} =-g&\left(| G\rangle\langle X_H|\hat{a}_H^\dagger+| X_H\rangle\langle B|\hat{a}_H^\dagger\right.
\\&\left.\hspace{0.25cm}+| G\rangle\langle X_V|\hat{a}_V^\dagger-| X_V\rangle\langle B|\hat{a}_V^\dagger\right)+H.c.\,.
\end{split}
\end{equation} 
Here, the four states of the QD are represented by the biexciton state $| B\rangle$, 
the two possible exciton states $| X_H\rangle$ and $| X_V\rangle$, and the ground state $| G\rangle$. 
The exciton states as well as the two photon modes are labeled with $H$ (horizontal polarization) and $V$ (vertical polarization).
The bosonic operator $\hat{a}_{H/V}^\dagger$ creates one photon with  frequency $\omega_{H/V}^c$
and corresponding polarization $H$ or $V$ and $H.c.$ denotes the Hermitian
conjugate. The light-matter coupling strength $g$ is assumed to be equal for all
couplings and the dipole approximation as well as the rotating-wave approximation
are used. The energies $\hbar\omega_{H/V}$ denote the exciton energies,
while the energy of the biexciton is $\hbar(2\bar{\omega}_{X}-\omega_B)$, where
$E_{B}=\hbar\omega_{B}$ represents the biexciton binding energy and 
$\hbar\bar{\omega}_{X}=\hbar(\omega_{H}+\omega_{V})/2$ is the mean exciton energy.
The energy of the ground state is set to zero. When
the QD is initially prepared in the biexciton state without any photons present
in the two orthogonal cavity modes, the total number of excitations (number of
excitons plus number of photons) is initially two. Since without losses the excitation number is conserved, the
number of states that are accessible by the coherent QD-cavity coupling is
restricted to five states of the form $|\chi,n_H,n_V\rangle$ with $\chi$ denoting the
QD state and $n_{H/V}$ the number of photons present in the corresponding cavity mode.
To be specific, these five states are given by $|B,0,0\rangle$, $|X_H,1,0\rangle$, $|X_V,0,1\rangle$,
$|G,2,0\rangle$, and $|G,0,2\rangle$.
States with lower excitation numbers become accessible via cavity losses
by removing photons from the system. However, we do not need to consider these states explicitly
in our calculations, first,  because the corresponding dynamical variables 
do not couple back to the dynamics of the above five states and, second,
only states with at least two photons contribute to the concurrence \cite{new_paper,Wootters:1998}, which
is the target quantity of our analysis.  

\begin{figure}
\centering
\includegraphics[width=0.75\columnwidth]{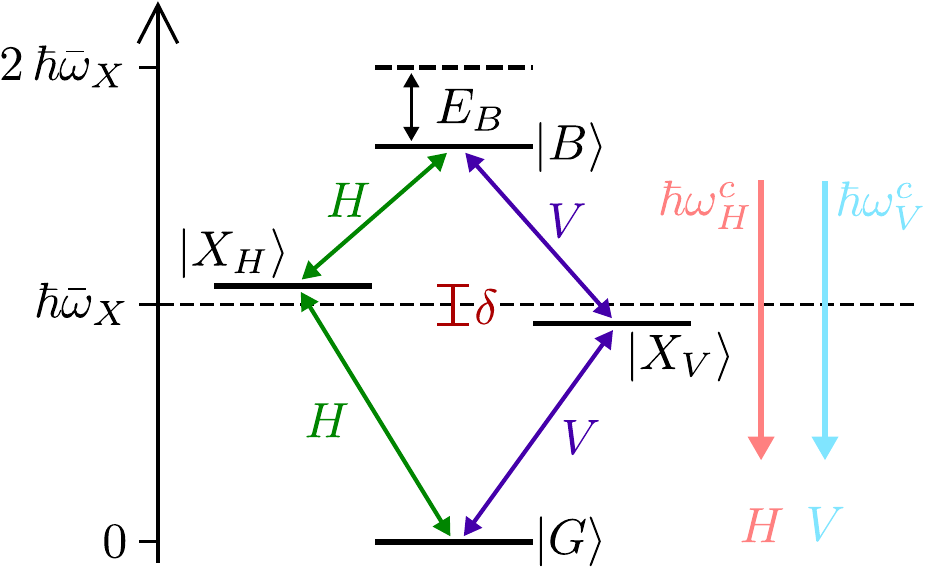}
\caption{Schematic sketch of the biexciton cascade with a fine-structure splitting $\delta$ between the two exciton states, a mean exciton energy 
$\hbar\bar{\omega}_{X}=\hbar(\omega_{H}+\omega_{V})/2$, and a possible biexciton binding energy $E_B$.
In general, both cavity modes can be detuned from the electronic transitions of the QD.}
\label{fig:model}
\end{figure}

In Fig.~\ref{fig:model} a schematic sketch of the biexciton
cascade with the two cavity modes is shown. Because of the exchange interaction
the two exciton states $X_H$ and $X_V$ are split by the fine-structure splitting
$\delta$ symmetric to the mean exciton energy
$\hbar\bar{\omega}_X$. Thus, the energy of
the horizontally polarized exciton state is
$\hbar\omega_H=\hbar\bar{\omega}_X+\delta/2$ and the energy of the vertically
polarized exciton state is $\hbar\omega_V=\hbar\bar{\omega}_X-\delta/2$. 
Furthermore, a possible biexciton binding energy $E_B$ can lower the energy of the biexciton state with
respect to $2\hbar\bar{\omega}_X$. 
In general, the energies of the two orthogonally polarized cavity modes do not match any of the
electronic transition energies of the QD.

In addition to the light-matter interaction also a pure dephasing type coupling to
a continuum of LA phonons is included in the model via
\begin{equation}
\label{eq:H_QD-phon}
\begin{split}
\hat{H}_\text{QD-phon}=&\sum_{\bf q}\hbar\omega_{\bf q}\hat{b}_{\bf q}^\dagger\hat{b}_{\bf q}
\\&+\sum_{{\bf q},\chi}n_\chi\left(\gamma_{\bf q}
\hat{b}_{\bf q}^\dagger+\gamma_{\bf q}^*\hat{b}_{\bf q}\right)| \chi\rangle\langle \chi|.
\end{split}
\end{equation} 
Here, $n_\chi$ denotes the number of excitons in the different QD states $|\chi\rangle$ and $\gamma_{\bf q}$ is the coupling
constant. We account for deformation potential coupling which is known to dominate for GaAs-type QDs \cite{dotspektren:02} and 
$\hat{b}_{\bf q}^\dagger$ are bosonic creation operators for phonons with energy $\hbar\omega_{\bf q}$ in the mode with wave vector ${\bf q}$.

Finally, possible cavity losses of photons are taken into account using the Lindblad operator
\begin{equation}
\label{eq:L_cav}
\mathcal{L}_\text{cav}\left[\hat{\rho}\right]=\sum_{\ell=H,V}\frac{\kappa_\ell}{2}\left(2\,\hat{a}_\ell\hat{\rho}\hat{a}_\ell^\dagger
-\hat{\rho}\hat{a}_\ell^\dagger\hat{a}_\ell-\hat{a}_\ell^\dagger
\hat{a}_\ell\hat{\rho}\right)
\end{equation}
which allows the inclusion of non-Hamiltonian dynamics while preserving the physically important properties of 
the statistical operator\cite{Lindblad:1976}. 
In the following we assume the loss rates for the two differently polarized cavity modes to be equal ($\kappa_H=\kappa_V=\kappa$).

Longitudinal optic (LO) phonons have been shown to affect the two-photon entanglement by multiphonon transitions to the continuum
of wetting layer states \cite{PhysRevB.81.195319}.
This mechanism is, however, negligible for temperatures below $\sim$ 80 K. Since all major findings of the present paper occur at
much lower temperatures, effects of LO phonons can safely be disregarded. Nevertheless, we show in the present paper a few results
for temperatures above 80 K in order to illustrate how the contribution of LA phonons behaves at elevated temperatures.

\subsection{Method}
\label{subsec:method}

Equation~(\ref{eq:dynamic_density_matrix}) is numerically solved by using a real-time path-integral (PI) approach.
As almost all modern implementations of the real-time PI concept, also our simulations are based on an iteration scheme for the so called {\em augmented density matrix} which was introduced 
in the pioneering work of Makri and Makarov \cite{doi:10.1063/1.469508,doi:10.1063/1.469509}.
This scheme  exploits the finiteness of the environment memory to obtain an efficient algorithm for performing efficiently a numerically complete summation over the paths.
A specialization of these general ideas to QDs with a super-Ohmic pure-dephasing coupling to a continuum of phonons has been worked out, e.g., in Ref.~\onlinecite{PhysRevB.83.094303}.
Important for the present investigations are two recent extensions of the standard PI treatment.
The first is a translation of the PI method from the usual Hilbert space formulation to Liouville space\cite{Andi:2016}.
In this way, the non-Hamiltonian contributions to the dynamics, like, e.g., Lindblad-type loss rates, can be accounted for in a natural way while still treating the phonons without approximation to the model.
The second is a reformulation of the PI algorithm such that now a partially summed augmented density matrix is iterated.
This reformulation is described in detail in the supplement of Ref.~\onlinecite{cavitypaper}, which in principle contains all information about the actually used PI method. For systems like QDs coupled to cavities the reformulation reduces the numerical demands by many orders of magnitude and thus numerically complete simulations for such systems would not be feasible without it.
The numerical efficiency might be further boosted by using recently developed tensor-network techniques \cite{Strathearn2018} which could further extend the applicability of PI methods in future work.

We consider a spherically symmetric GaAs QD with a harmonic oscillator confinement 
resulting in an electron (hole) confinement length $a_e=3$ nm ($a_h=a_e/1.15$). The deformation potential
constants and the mass density as well as the sound velocity are taken from the literature\cite{Krummheuer:2005}
and enter the phonon spectral density
\begin{align}
J_{\chi\chi'}(\omega)&= n_{\chi}\, n_{\chi'}\,J(\omega)\\
\intertext{with}
J(\omega) &= \sum_{\bf q}\gamma_{\bf q}\gamma_{\bf q}^{*}\delta(\omega-\omega_{\bf q})
\label{eq:spectral_density_general}
\end{align}
appearing in the memory kernels of the PI approach\cite{Andi:2016}. 
It is worthwhile to note that all phonon related influences on the dynamics of the QD-photon system enter only
via $J(\omega)$. Thus the assumption of a symmetric QD does not entail a loss of generality as long as only the
dynamics of QD and photons are concerned, since as shown in Ref.~\onlinecite{Lueker2017}
for any QD (not necessarily assuming a symmetric confinement) 
it is always possible to find a symmetric QD with the same $J(\omega)$.

\begin{figure}
\centering
\includegraphics[width = \columnwidth]{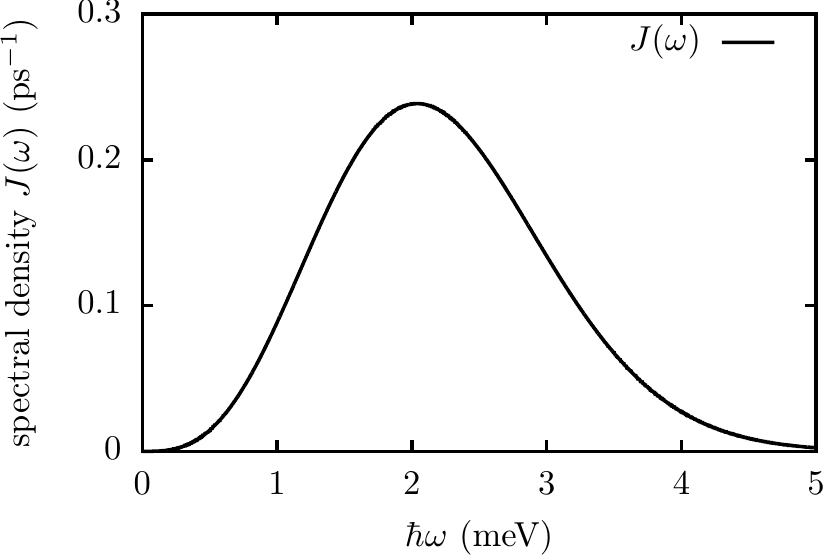}
\caption{Phonon spectral density $J(\omega)$ for a spherical GaAs QD with 
an electron (hole) geometrical confinement length $a_e=3$ nm ($a_h=a_e/1.15$). 
The deformation potential constants and the mass density, as well as the sound velocity 
for a GaAs QD are taken from Ref.~\onlinecite{Krummheuer:2005} and are listed
in Table~\ref{tab:QD_Parameter}. An explicit formula for $J(\omega)$ can
be found in Ref.~\onlinecite{PhysRevB.83.094303} or Ref.~\onlinecite{Andi:2016}.}
\label{fig:spectral_density}
\end{figure}

In Fig.~\ref{fig:spectral_density}  $J(\omega)$ is
depicted for the chosen parameters of the QD. For low frequencies $J(\omega)$ 
approaches zero $\sim\omega^{3}$ as can be seen from the explicit
expression for the deformation potential coupling \cite{dotspektren:02}.
We are therefore dealing with a coupling of super-Ohmic type
which is responsible for striking non-Markovian effects such as 
the nonexponential partial loss of coherence\cite{dotspektren:02,besombes:01}.
Furthermore, we note a pronounced maximum at about 2~meV that is the origin
of the resonant  structure of the phononic response \cite{machnikowski:04a,RabiRevival}.

Assuming that initially the phonons are in thermal equilibrium and the electronic system is prepared  
in the biexciton state without photons, our PI approach delivers 
the time dependence of the reduced density matrix $\hat{\bar{\rho}}$ 
in the subspace spanned by the five states $| B,0,0\rangle$, $| X_H,1,0\rangle$, $| X_V,0,1\rangle$, 
$| G,2,0\rangle$, and $| G,0,2\rangle$, where the phonon degrees of freedom have been traced out.

\subsection{Concurrence}
\label{subsec:concurrence}

As a measure for the degree of entanglement we use the
concurrence of simultaneously emitted photons that for brevity will be referred to 
in the following simply as the {\em concurrence}.
This quantity can be
directly calculated from the time-averaged values of the reduced density matrix
$\hat{\bar{\rho}}$ of the
system\cite{new_paper,PhysRevB.81.195319,carmele2011}. The time-dependent
populations of the two states where two photons are present and the coherences
between these states are given by
\begin{equation}
\label{eq:occupations_and_coherences}
\rho_{mn}(t) = \langle mm|\hat{\bar{\rho}}(t)| nn\rangle
\end{equation} 
with $m,n \in \{H,V\}$.
Here $| HH\rangle := | G,2,0\rangle$ is the state with two horizontally polarized photons and $| VV\rangle := | G,0,2\rangle$
denotes the state with two vertically polarized photons. 
The corresponding time-averaged quantities $\bar{\rho}_{mn}$ are calculated according to
\begin{equation}
\label{eq:rho_averaged}
\bar{\rho}_{mn} = \frac{1}{T_\text{av}}\int_0^{T_\text{av}} \rho_{mn}(t)\,\mathrm{d}t.
\end{equation}
From these quantities, the concurrence $C$ is derived as \cite{new_paper,Wootters:1998}
\begin{equation}
\label{eq:def_concurrence_carmele}
C = 2\,|\bar{\rho}_{HV}^N|,
\end{equation} 
where all quantities entering the normalized two-photon coherence
\begin{equation}
\label{eq:normalized_two-photon-coherence}
\bar{\rho}_{HV}^N = \frac{\bar{\rho}_{HV}}{\bar{\rho}_{HH}+\bar{\rho}_{VV}}
\end{equation} 
are evaluated in the limit $T_\text{av}\rightarrow\infty$.
We average the time-dependent quantities $\rho_{mn}(t)$ until all
excitations have left the cavity and the system has reached its ground state $|
G, 0, 0\rangle$. 
Experimentally, the concurrence $C$ is accessible\cite{PhysRevLett.101.170501,Bounouar18} by measuring the two-photon
correlation function $G_{ij,kl}^{(2)}(t,\tau)$ and extrapolating towards zero delay time $\tau = 0$.

Before presenting the results of our calculations, let us briefly comment on different measures to quantify the entanglement in the biexciton cascade and the impact of the cavity loss rate (a more extended discussion of these issues can be found, e.g., in Ref.~\onlinecite{new_paper}).
Indeed, for an analysis of polarization entanglement there is a variety of choices for selecting photon pairs for which to calculate the concurrence.
Probably the most widely used choice is to inspect the concurrence of all photon pairs that are  detected in coincidence measurements without discriminating between the detection times of the two photons.
The obvious advantage of this scheme is the high signal yield.
For the corresponding theoretical description, the calculation of the two-time two-photon correlation function $G^{(2)}(t,\tau)$  is required\cite{Pfanner2008,Schumacher:2012,heinze2017,new_paper,multi-time}.
Another approach is to consider the concurrence of frequency filtered coincidence measurements \cite{EdV,Pfanner2008} or for a subsystem of the detected photons, e.g., by only selecting photon pairs with equal emission times from the cavity\cite{carmele2011,EdV}.
As stated previously, we follow the latter scheme and focus on the concurrence of simultaneously emitted photon pairs. The reasons behind this choice are presented in the following.

The concurrence calculated for a selected subset of photons is in general quantitatively as well as qualitatively different from the concurrence obtained for another photon subset.
For example, it has been found \cite{new_paper} that the concurrence of simultaneously emitted photons  shows qualitatively different trends with varying cavity losses than observed for photon pairs without selection of the emission time (an increase of the concurrence with rising loss rate is turned into a decrease).
Thus these two concurrences calculated for different photon subsets cannot be equivalent measures for the same physical quantity.
Nevertheless, in both cases, the phonon impact is reduced with rising loss rate and the concurrence approaches its phonon-free value in the limit of infinite losses \cite{new_paper}.
This can be explained by noting that the phonon impact requires a finite time to develop. The loss rate limits the available time window and when the latter becomes too small the phonons cannot efficiently act on the QD degrees of freedom.
Besides the different trends regarding the cavity loss rate, experiments\cite{PhysRevLett.101.170501,Bounouar18} as well as theory \cite{EdV,new_paper} indicate that simultaneously emitted photons exhibit a significantly higher degree of entanglement and are much less affected by the which-path information introduced by a finite fine-structure splitting than photon pairs detected without emission time selection. In particular, the concurrence of simultaneously emitted photons represents an upper limit for the achievable degree of entanglement in the latter situation.
Since we are interested in the highest possible degree of entanglement for a given QD-cavity configuration we solely concentrate on the concurrence of simultaneously emitted photons. 
Detecting photon pairs without emission time filtering, on the other hand, would maximize the emission efficiency.

\begin{figure*}
\centering
\includegraphics[width=\textwidth]{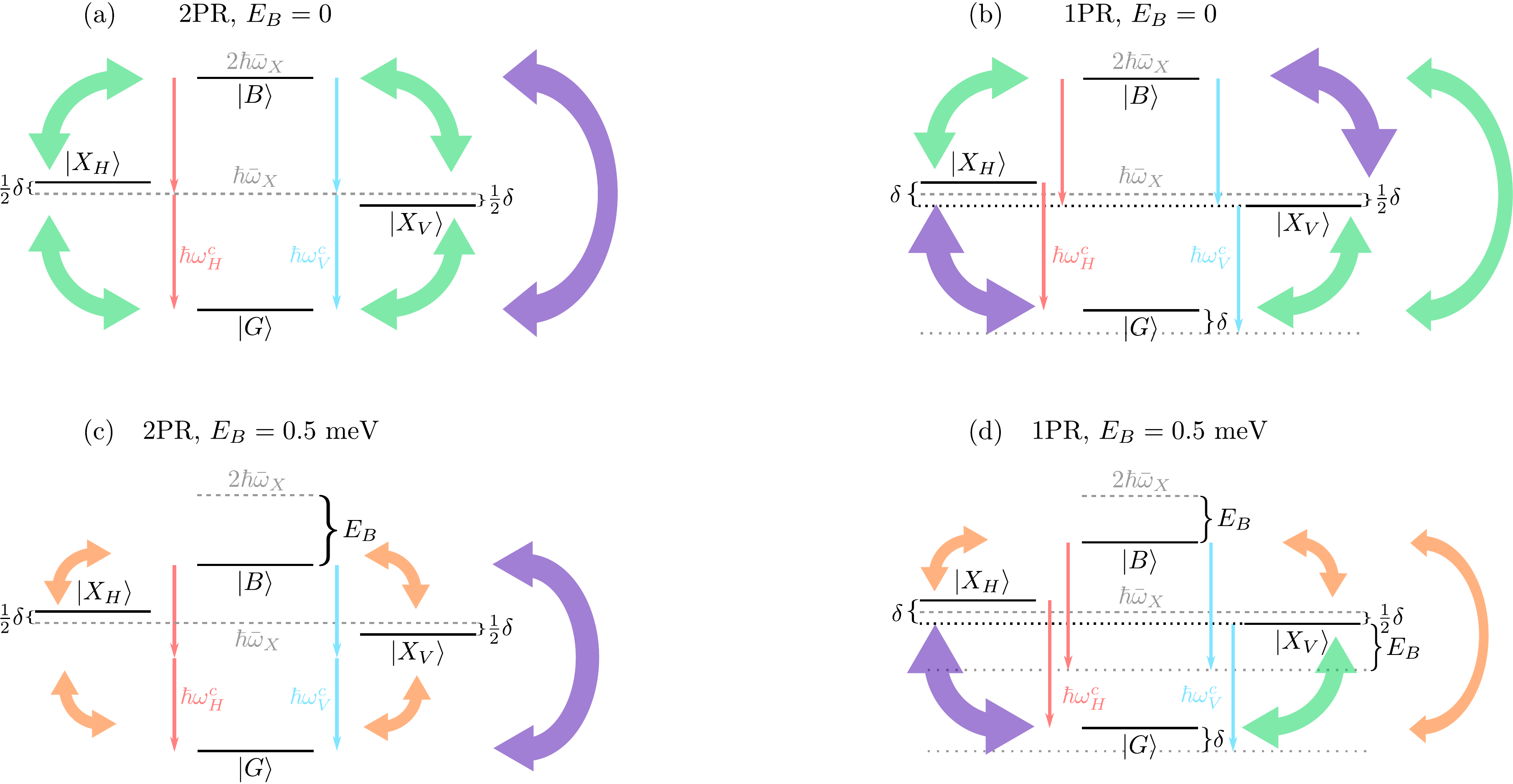}
\caption{Schematic sketches of the four different configurations of the QD-cavity
system studied in this paper. Big curved purple arrows indicate transitions
that are resonant with the corresponding cavity mode. Transitions which are
detuned on the order of the fine-structure splitting $\delta$ are represented by medium-sized
curved green arrows and small curved orange arrows indicate transitions where
the detuning is on the order of the biexciton binding energy $E_B$.}
\label{fig:schematic}
\end{figure*}

\section{Different cavity configurations} 
\label{sec:configurations}

In this section, four configurations of the QD-cavity system are considered which differ in the value of the biexciton binding energy $E_B$ as well as
in the way the cavity modes are energetically positioned.
Throughout this paper, the two orthogonal linearly polarized cavity modes are assumed to have the same frequency ($\omega_H^c=\omega_V^c$).
The main difference between the configurations is the position of the cavity modes with respect to the QD transitions.
When the cavity modes are kept in resonance with the direct two-photon transition to the biexciton, such that
$\omega_{H/V}^c=\bar{\omega}_{X}-\omega_B/2$, 
we refer to the configuration as \emph{two-photon resonant} (2PR).
In contrast, if the cavity mode frequencies are chosen to match the transition frequency of one of the excitons (without loss of generality we choose
the $H$ exciton),
such that $\omega_{H/V}^c=\omega_H$, we refer to the configuration as \emph{one-photon resonant} (1PR).
In both configurations, we further distinguish between the case of a vanishing biexciton binding energy and
a finite value  of the latter (in this paper, we consider finite values $0.5\, \text{meV}\le  E_{B}\le 6$ meV).
Note that if a finite biexciton binding energy is introduced, the energy
of the biexciton state is no longer the sum of the energies of the two exciton states.

In Fig.~\ref{fig:schematic} schematic sketches of the 2PR and 1PR configuration
with and without a biexciton binding energy are shown.
In order to highlight the difference concerning the respective resonance situations,
QD transitions are marked by three types of curved arrows in the figure that correspond to
different detunings of these transitions from the cavity mode frequency (red or blue straight arrow). 
Resonant transitions are represented by big curved purple arrows. 
The medium-sized curved green arrows denote transitions which are detuned on the order of the fine-structure splitting $\delta$
and strongly off-resonant transitions with a detuning on the order of the biexciton binding energy $E_{B}$
(typically much larger than $\delta$) are indicated by small curved orange arrows.

The special characteristic of the 2PR configurations [Figs.~\ref{fig:schematic}(a) and \ref{fig:schematic}(c)] is that the $|G\rangle \leftrightarrow |B\rangle$
transition is resonant with a direct two-photon emission or absorption process, respectively. 
Therefore, there are two competing channels for the biexciton decay. 
The biexciton state can decay either via two sequential single-photon emission processes via the exciton states or via a 
coherent two-photon process from the biexciton state directly to the ground state.
For vanishing $E_B$  [Fig.~\ref{fig:schematic}(a)] the energies
of the exciton states  $\hbar\omega_{H/V}=\hbar\bar{\omega}_X\pm\delta/2$ are detuned  by $\pm\delta$/2 from the cavity modes 
which are fixed at $\omega_{H/V}^c=\bar{\omega}_X$. 
Thus all four electronic transitions involved in the sequential emission paths  are weakly detuned by
half the value of the fine-structure splitting $\delta$.  
The direct two-photon processes stay resonant in the 2PR configuration when a finite binding energy $E_B$ is introduced as the cavity
modes are changed accordingly. 
But the four electronic transitions involving an exciton state become strongly detuned on the order of half the biexciton binding
energy $E_B/2$ [Fig.~\ref{fig:schematic}(c)] when $E_B$ is finite.  

In the 1PR configurations [Figs.~\ref{fig:schematic}(b) and \ref{fig:schematic}(d)], the $|X_H\rangle \leftrightarrow |G\rangle$ transition is chosen to be resonant with 
the corresponding cavity mode.
Therefore, in the case of a vanishing biexciton binding energy [Fig.~\ref{fig:schematic}(b)], the $|X_V\rangle \leftrightarrow |B\rangle$ transition is also resonant, 
whereas the two remaining cascade transitions as well as the direct two-photon processes are detuned by the value of the splitting $\delta$.
Introducing a finite biexciton binding energy does not change the situation for the exciton-to-ground-state transitions but the two transitions between 
the biexciton state and one of the exciton states as well as the direct two-photon processes are now
strongly off resonant and detuned on the order of $E_B$ [cf. Fig.~\ref{fig:schematic}(d)]. 

\section{Results} 
\label{sec:results}

In this section we analyze how the degree of entanglement between the two states with two photons  ($| HH\rangle$ and $| VV\rangle$)
is affected by various system parameters. 
As mentioned before, the system is initially prepared in the biexciton state without any photons and the phonons are assumed to be initially 
in thermal equilibrium. 
If not stated otherwise, a light-matter coupling strength $g=0.1$~meV, a finite exciton splitting $\delta=0.1$~meV,
a biexciton binding energy $E_{B}= 0.5$ meV, 
and a cavity loss rate $\kappa=0.025$~$\text{ps}^{-1}$ corresponding to a cavity quality factor $Q\approx 45\,000$ are used.
Table~\ref{tab:QD_Parameter} displays these default values and all other material parameters used for the numerical calculations.
The given value for the biexciton binding energy $E_B$ is the difference between twice the polaron shifted mean exciton energy and the polaron shifted biexciton energy. In the corresponding phonon-free situation the value for $E_B$ is kept the same in order to compare QD-cavity systems with identical energetic detunings between the cavity modes and the QD transition energies.
After quantifying the competition between direct two-photon and sequential single-photon processes in Sec.~\ref{subsec:competition}, the dependence of the concurrence on the
exciton fine-structure splitting is investigated in Sec.~\ref{subsec:splitting}.
Finally, we discuss the temperature dependence of the concurrence for fixed splittings in Sec.~\ref{subsec:temperature}.

\begin{table}
\centering
\caption{Material parameters for the GaAs quantum dot and the default values 
  for the system parameters: light-matter coupling strength $g$, exciton splitting $\delta$,
  biexciton binding energy $E_B$, and cavity loss rate $\kappa$. If not stated otherwise these default values are used for the calculations.}
\label{tab:QD_Parameter}
\begin{tabular}{l c c}
\hline\hline
Parameter & & Value\\
\hline
Electron geometrical confinement length (nm) & $a_e$ & 3.0 \\
Hole geometrical confinement length (nm) & $a_h$ & $a_e/1.15$ \\
Mass density (kg/$\mathrm{m^3}$) & $\rho$ & 5370\cite{Physics_of_group_IV_elements} \\
Longitudinal sound velocity (m/s) & $c_s$ & 5110\cite{Physics_of_group_IV_elements} \\
Electron deformation potential constant (eV) & $D_e$ & 7.0\cite{PhysRevB.54.4660} \\
Hole deformation potential constant (eV) & $D_h$ & -3.5\cite{PhysRevB.54.4660} \\
Light-matter coupling strength (meV) & $g$ & 0.1 \\
Exciton fine-structure splitting (meV) & $\delta$ & 0.1 \\
Biexciton binding energy (meV) & $E_B$ & 0.5 \\
Cavity loss rate ($\mathrm{ps^{-1}}$) & $\kappa$ & 0.025\\
\hline\hline
\end{tabular}
\end{table}

\subsection{Competition between direct two-photon and sequential single-photon transitions}
\label{subsec:competition}

As pointed out before, e.g., by Schumacher et al.\cite{Schumacher:2012} or del Valle \cite{EdV},
the competition between the direct two-photon processes from the biexciton state to the ground state and the sequential 
single-photon processes via the exciton states is of utmost importance for the concurrence.
Obviously, by considering configurations with different resonance situations, in particular when 
switching between 2PR and 1PR configurations, we are comparing situations with different relative
importance of two-photon and sequential single-photon processes. 
To quantify the relative impact of these processes, we introduce the quantity
\begin{equation}
\label{eq:ratio}
r_{2P/1P}=\frac{|\bar{\rho}_{B,HH}|+|\bar{\rho}_{B,VV}|}{\bar{\rho}_{X_H}+\bar{\rho}_{X_V}},
\end{equation} 
i.e., $r_{2P/1P}$ is a ratio where the numerator is derived from  the coherences $\rho_{B,HH}=\langle B,0,0|\hat{\bar{\rho}}| G,2,0\rangle$ and 
$\rho_{B,VV}=\langle B,0,0|\hat{\bar{\rho}}| G,0,2\rangle$ between the biexciton and the ground state with either two horizontally or vertically polarized photons.
The denominator represents the total exciton occupation $\rho_{X_H}+\rho_{X_V}$, where $\rho_{X_{{H}}}$ and $\rho_{X_{{V}}}$
denote the occupations of the states $| X_H,1,0\rangle$ and $| X_V,0,1\rangle$, respectively.
The bar over these quantities indicates a time averaging as introduced in Eq.~(\ref{eq:rho_averaged}).

The coherences between the biexciton state and the two states containing the ground
state represent a measure for the direct two-photon processes. 
Note that by inspecting the equations of motion for all elements of the reduced density matrix it becomes apparent that 
only the equations for these coherences introduce a resonance when the 
biexciton-to-ground-state transition frequency matches twice the photon frequency. 
This is the distinctive property of a two-photon process. 
In contrast, the characteristic feature 
of sequential single-photon emission processes is the occupation of the intermediate electronic states,
in our case the excitons. Thus the total exciton occupation reflects
the importance of sequential processes.
Altogether, this justifies that the ratio $r_{2P/1P}$ is a possible measure for the relative importance of the direct two-photon processes compared with 
the sequential single-photon processes.

\begin{figure}
\centering
\includegraphics[width = \columnwidth]{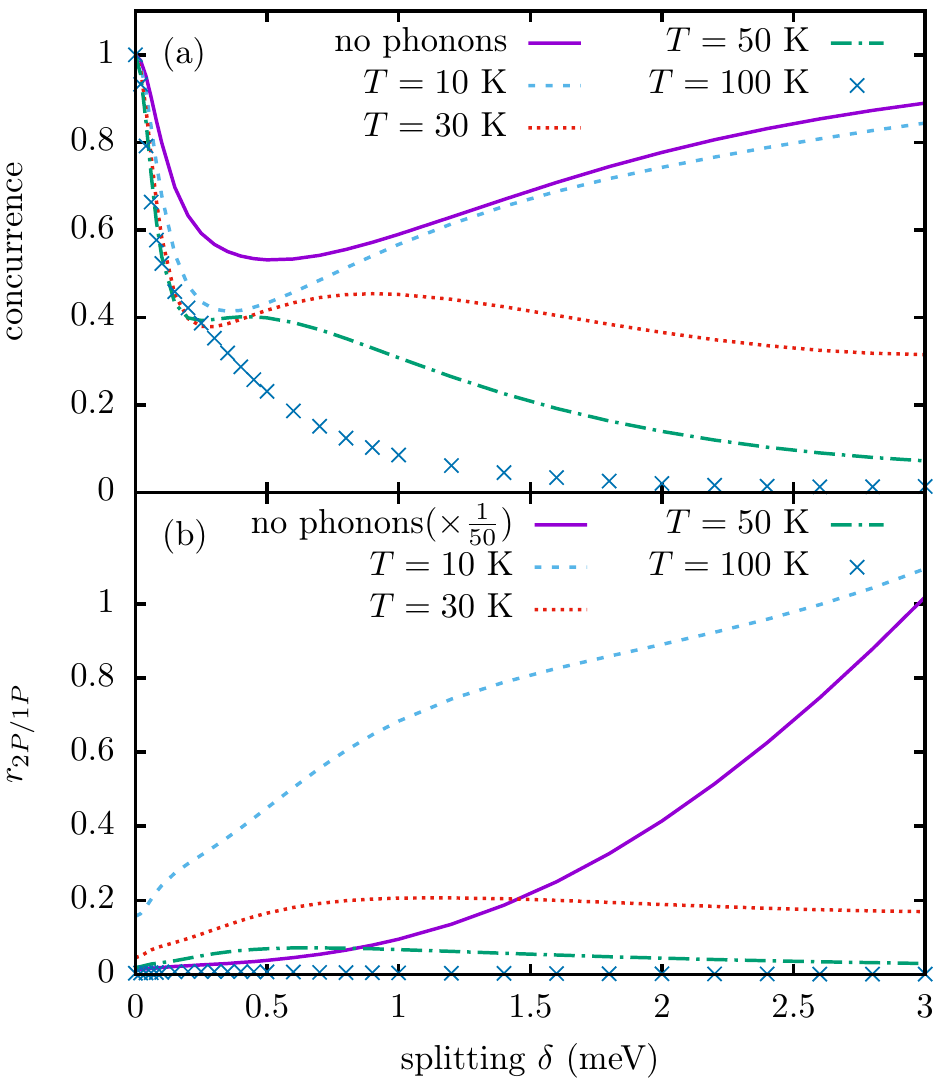}
\caption{Comparison of the concurrence [panel (a)] and a measure for the relative
importance of two-photon and sequential single-photon processes $r_{2P/1P}$ [panel (b)] as a function of
the exciton splitting $\delta$. Different temperatures as well as the limit without phonons
are considered. 
Note that $r_{2P/1P}$ is scaled by the factor 1/50 in the phonon-free case. 
The cavity modes are arranged in the 2PR configuration and a vanishing biexciton binding energy is assumed.}
\label{fig:2PR_analyse}
\end{figure}

Figure~\ref{fig:2PR_analyse} displays the concurrence [panel (a)] along with $r_{2P/1P}$ [panel (b)] as a function of $\delta$.
The analysis is carried out exemplarily for the 2PR configuration with $E_B = 0$ for four temperatures as well as for the phonon-free case.
The plotted range for the fine-structure splitting is chosen larger than usually covered by typical QDs 
as the role of the two-photon processes can be better highlighted on this extended scale.

As can be seen in Fig.~\ref{fig:2PR_analyse}(a), the concurrence exhibits a nonmonotonic
dependence on the exciton splitting for low temperatures and the phonon-free situation. 
This behavior can be traced back to the competition between two-photon and sequential single-photon processes.
Recalling that in the 2PR configuration the two-photon processes are chosen to be always resonant independent of $\delta$,
it follows that any which-path information introduced by the fine-structure splitting affects only the
sequential single-photon processes.
Figure~\ref{fig:2PR_analyse}(b) reveals a dominance of sequential emission processes for small exciton splittings.
Therefore, the concurrence decreases with rising $|\delta|$ in the small splitting limit since 
(i) the which-path information is larger for larger $|\delta|$ and 
(ii) it efficiently affects the concurrence due to the dominance of sequential single-photon processes.

As the splitting increases further, $r_{2P/1P}$ rises because the single-photon processes become more off-resonant and thus the 
relative importance of two-photon processes grows, since the latter are always resonant.  
When either the interaction with the phonons is switched off or the temperature is low enough, 
$r_{2P/1P}$ increases strongly for larger exciton splittings, indicating a dominance of two-photon processes [cf. Fig.~\ref{fig:2PR_analyse}(b) for $T = 10$~K]. 
As a result, the concurrence rises and eventually approaches unity because the which-path information introduced by the
exciton splitting is no longer tested.
The local maximum of the concurrence seen at higher temperatures of 30~K and 50~K can also be understood with the help of $r_{2P/1P}$ since it 
shows a similar behavior. 
Hence the nonmonotonic behavior of the concurrence is a result of the competition between the coherent direct two-photon and sequential single-photon emission
processes.  

At higher temperatures the relative importance of the sequential emission processes is raised, as can be seen in Fig.~\ref{fig:2PR_analyse}(b). 
As the electronic transitions become detuned from the corresponding cavity modes the sequential single-photon processes are assisted by phonon absorption and emission
processes to compensate the energy differences, an effect which is enhanced with increasing temperature.
In addition, for larger exciton splittings the phonon spectral density is effectively probed at higher values of $\omega$ (on the order of $\delta$),
resulting in a stronger phonon influence on the system (cf.~Fig.~\ref{fig:spectral_density}). 
Furthermore, coherences, such as the ones relevant for the two-photon processes, are more strongly affected by phonon-induced decoherence than occupations. 
The combination of these effects explains why $r_{2P/1P}$ decreases for larger splittings at higher temperatures and the concurrence approaches zero.

\subsection{Dependence of the concurrence on the exciton splitting for different configuratons}
\label{subsec:splitting}
 
\begin{figure*}
\centering
\includegraphics[width = \textwidth]{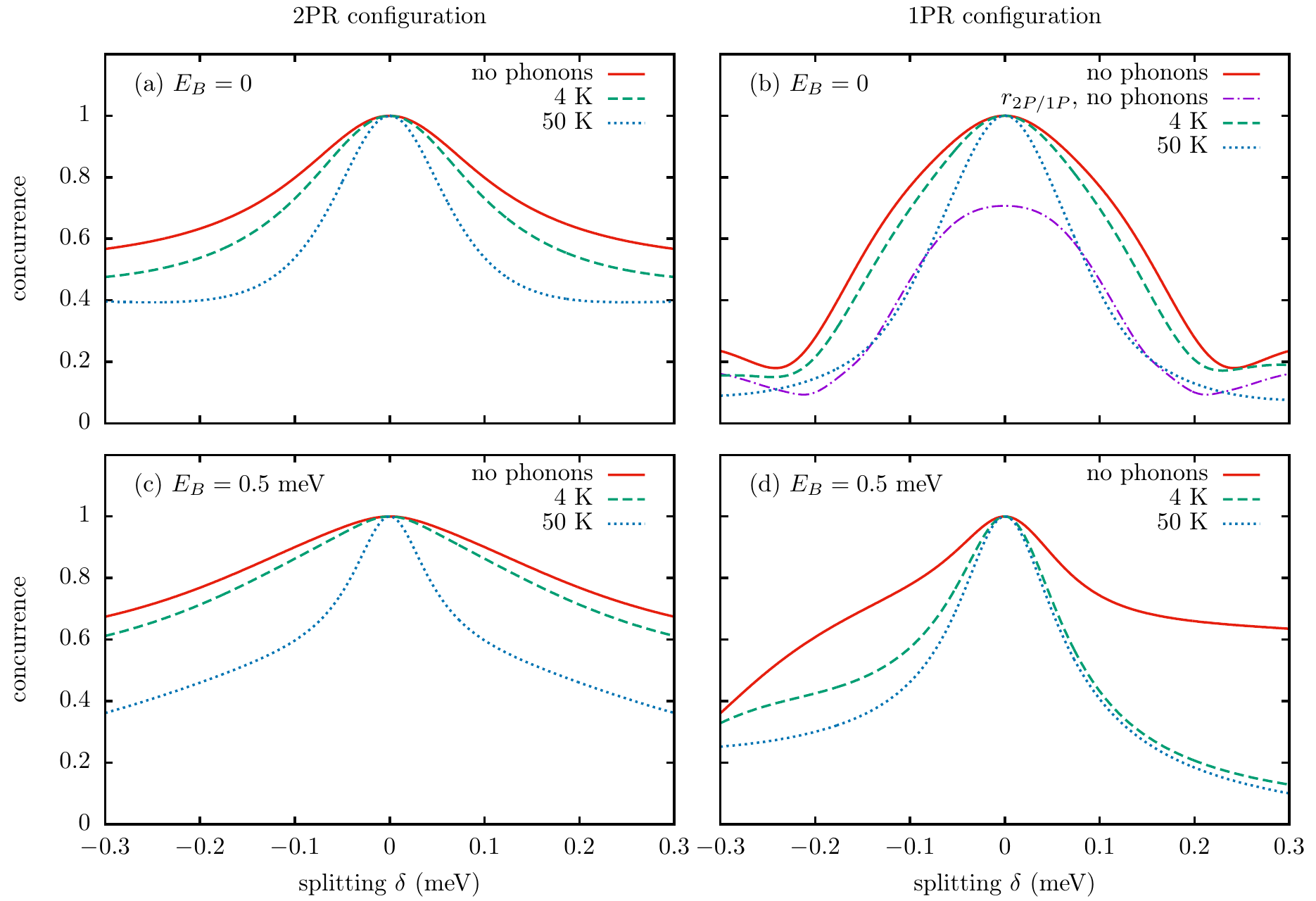}
\caption{Impact of the exciton splitting $\delta$ on the concurrence for all
four cavity configurations introduced in Sec.~\ref{sec:configurations} and
depicted in Fig.~\ref{fig:schematic}. Results are shown for two different
temperatures and without phonons. 
Panel (b) additionally displays the ratio $r_{2P/1P}$ for the phonon-free case.}
\label{fig:all_config_delta}
\end{figure*}

For all four QD-cavity configurations illustrated in Fig.~\ref{fig:schematic}, the dependence of the concurrence on the exciton splitting is shown in 
Fig.~\ref{fig:all_config_delta}.
First of all, for a vanishing fine-structure splitting, the concurrence is strictly one regardless of the phonon influence since the system is completely symmetric with respect to $H \leftrightarrow V$ so that no which-path information is introduced.
This result was also found on the basis of a phenomenological rate equation approach for the phonon-induced pure dephasing \cite{carmele2011}.

With increasing $|\delta|$ the concurrence decreases, reflecting the increase of which-path information. 
Furthermore, phonons generally reduce the concurrence for a given splitting, an effect which typically becomes more pronounced at higher temperatures. 
This can be understood by noting that phonons typically enhance the differences between different pathways and thus increase the which-path information.
To see this, we first recall that, when the electronic transitions of the QD are detuned
from the corresponding cavity modes, the photon emission processes are assisted by phonon emission and absorption processes to compensate 
the energy differences. 
For a finite splitting, depending on the configuration, the two sequential emission paths differ either in the values or the order of the
detunings and are therefore influenced differently by the phonons. 
For example, in the 2PR configuration with $\delta>0$ and $E_B=0$, the sequential emission process of two horizontally polarized photons is at first
assisted by phonon absorption and in the second step phonon emission occurs for the exciton-to-ground-state transition. 
This order is reversed for the sequential emission path for two vertically polarized photons. Obviously, this enhances the difference
between both pathways compared with the phonon-free case at least at low temperatures where emission and absorption are noticeably different.
In general, as discussed in Sec.~\ref{subsec:competition}, with increasing temperature more phonons are thermally
excited and sequential one-photon transitions can
be more efficiently bridged by phonon-assisted processes.
In combination with the  increased phonon-induced decoherence, this leads to a smaller impact of two-photon transitions and therefore
a lower concurrence at higher temperatures.

Despite these common tendencies, the detailed dependences of the concurrence on the exciton splitting differ significantly in the respective configurations. 
For $E_B=0$, the results for the 2PR configuration [Fig.~\ref{fig:all_config_delta}(a)] and the 1PR configuration
[Fig.~\ref{fig:all_config_delta}(b)] are qualitatively similar for small $|\delta|$ but differ strongly for larger detunings. 
This can be understood by consulting Fig.~\ref{fig:schematic}(a) and Fig.~\ref{fig:schematic}(b) which reveals that 
these configurations become identical in the limit of vanishing splitting.
The corresponding concurrences are thus very close to each other for small exciton splittings.

The deviation for larger splittings between the two configurations can be explained by the competition
between the coherent direct two-photon and the sequential single-photon processes. 
In the 2PR configuration, the relative importance of two-photon processes rises with
increasing $|\delta|$ as already discussed in Sec.~\ref{subsec:competition}. 
However, compared with the 2PR case, the influence of two-photon processes is
reduced in the 1PR configuration since they are detuned by the exciton splitting 
[cf.~Fig.~\ref{fig:schematic}(b)].
Thus the concurrence in the 2PR configuration is significantly higher for larger $|\delta|$ than in the 1PR configuration.

Nevertheless, the competition between two-photon and single-photon processes also influences the 1PR configuration
where the concurrence exhibits a local minimum for low temperatures as well as without phonons, which means that this 
is not a phonon-induced effect. 
In fact, phonons cause this minimum to eventually disappear, as can be seen in Fig.~\ref{fig:all_config_delta}(b) at 50~K.
Figure~\ref{fig:all_config_delta}(b) reveals that the nonmonotonic behavior of the concurrence reflects the behavior of $r_{2P/1P}$.
Compared with the 2PR configuration [cf.~Fig.~\ref{fig:2PR_analyse}(a)], here the local minima are found already
at smaller $|\delta|$ because the electronic transitions of the QD are now detuned by the value of $\delta$, whereas the
detuning is only $\delta/2$ in the 2PR configuration.
Furthermore, although the value of $r_{2P/1P}$ at vanishing splitting without phonons in Fig.~\ref{fig:all_config_delta}(b)
is the same as in Fig.~\ref{fig:2PR_analyse}(b)
(note the different scaling in the latter figure), the ratio between two- and one-photon
processes is a decreasing function of $\delta$ in the 1PR configuration since a finite splitting
also causes a detuning of the direct two-photon transitions in this case.
This is in contrast to the 2PR case in Fig.~\ref{fig:2PR_analyse}, where $r_{2P/1P}$ rises with increasing $\delta$ in the phonon-free case.

Next, we consider the results for the 2PR configuration with a finite biexciton binding energy $E_B = 0.5$~meV
plotted in Fig.~\ref{fig:all_config_delta}(c).
In the phonon-free situation and for a temperature of 4~K the concurrence decreases only weakly with increasing $|\delta|$, 
but at $T=50$~K it is drastically reduced at finite $|\delta|$. 
Therefore, a finite biexciton binding energy in the 2PR configuration significantly affects the concurrence [cf. Fig.~\ref{fig:all_config_delta}(a)] since
it leads to strongly detuned biexciton-to-exciton and exciton-to-ground-state transitions, while the direct two-photon processes remain resonant. 
Thus, without phonons, the direct two-photon processes are strongly enhanced compared with the sequential single-photon processes, 
resulting in a significantly higher concurrence that is much less influenced by the splitting.

On the other hand, in the 2PR configuration with finite $E_B$, phonons with energies $\hbar\omega_{{\bf q}}\simeq E_{B}/2$
are required to
bridge the detunings of the sequential transitions.
In contrast, for vanishing $E_B$, the required phonon energies are given by the much smaller value of $|\delta|/2$.
At the same time, the relative weight of the phonon influence is proportional to $J(\omega)$. 
Figure~\ref{fig:spectral_density} shows that $J(\frac{E_{B}}{2\hbar}) > J(\frac{\delta}{2\hbar})$
for $E_B=0.5$~meV and $|\delta|<0.3$~meV,
i.e., the phonon influence and thus the temperature dependence of the concurrence is stronger for a finite biexciton binding energy. 
This results in the significantly larger difference of the concurrence for 4~K and 50~K in Fig.~\ref{fig:all_config_delta}(c) compared with curves for the
same parameters but vanishing binding energy in Fig.~\ref{fig:all_config_delta}(a).
We note in passing that, keeping the splitting in the typical range $|\delta|<0.3$~meV, for rather high values of the biexciton binding energy 
the relation $J(\frac{E_{B}}{2\hbar}) > J(\frac{\delta}{2\hbar})$ is reversed (cf.~Fig.~\ref{fig:spectral_density}). 
However, this limit is usually not reached since typical biexciton binding energies stay below $\sim 6$~meV.

In contrast to the configurations discussed up to now, the interaction with phonons in the 1PR configuration with a finite biexciton binding energy, 
depicted in Fig.~\ref{fig:all_config_delta}(d), drastically reduces the concurrence already at low temperatures. 
In this situation, both biexciton-to-exciton-transitions are strongly detuned from the corresponding cavity modes. 
The horizontally polarized exciton-to-ground-state transition is resonant by definition, while the vertically polarized one is
detuned by $\delta$. 
In addition, also the direct two-photon processes are highly off resonant. 
As all possible electronic transitions starting from the biexciton state are strongly detuned, the initially prepared occupation
of the biexciton state decreases only very slowly when phonons are not accounted for. 
Hence the occupations of the exciton states and the QD ground state with two photons are always very small.
In both the $H$ and $V$ pathway the exciton can be reached by emission of a photon only when a phonon with an energy on the
order of $\simeq E_{B}$ is absorbed. At this energy $J(\omega)$ is even
larger than in the 2PR configuration with finite $E_{B}$
where phonons with energies $\simeq E_{B}/2$ are required, which explains the dramatic drop of the concurrence from the phonon-free
case to the values obtained for 4K.
Furthermore, the concurrence is clearly asymmetric with respect to the exciton splitting in this configuration.
Especially in the phonon-free case the concurrence decays much stronger with rising $|\delta|$  for negative than for positive $\delta$.
This is due to the fact that for negative $\delta$ one comes closer to the condition that the transition from the biexciton to the
$H$ exciton is getting in resonance. Since the decay from the $H$ exciton to the ground state is held in resonance
in this configuration, the pathway
$|B\rangle\to|X_{H},1,0\rangle\to|G,2,0\rangle$ is strongly favored compared with
$|B\rangle\to|X_{V},0,1\rangle\to|G,0,2\rangle$, resulting in low values of the concurrence.
Interestingly, for $\delta>0$ the concurrence decreases only very little when the temperature is raised further from 4~K to 50~K. 

In general, the symmetry with respect to $\delta$ is found to be another distinguishing feature between the 2PR and the 1PR configuration. 
In the 2PR configuration, independent of the biexciton binding energy, the concurrence is a symmetric function of the splitting no matter whether 
phonons are included or not.
In contrast, in the 1PR configuration with a finite binding energy, the concurrence always shows an asymmetric dependence on $\delta$.
In this situation, changing the sign of $\delta$ changes the absolute value of the detuning between the horizontally polarized cavity mode
and the corresponding biexciton-to-exciton transition $\Delta_{B,X_H}(\delta) = E_B + \delta$, while the
absolute values of the detunings of the remaining 
sequential transitions are unaffected.
The direct two-photon processes are also detuned by the same value $\Delta E_{2P}(\delta) = E_B + \delta$. 
Since without phonons the dynamics depends only on the absolute values of the detunings between the electronic
transitions and their corresponding cavity modes, 
an asymmetric concurrence is expected, which is also visible when phonons are accounted for.
This asymmetry  is stronger at low temperatures since there phonon absorption and emission processes are not equally likely.
Turning finally to the 1PR configuration with $E_B=0$, changing the sign of the exciton splitting no
longer changes the absolute values of the detunings.
Thus, without phonons, the concurrence is once more symmetric and only a slight asymmetry is observed when phonons are included.

\begin{figure*}
\centering
\includegraphics[width = \textwidth]{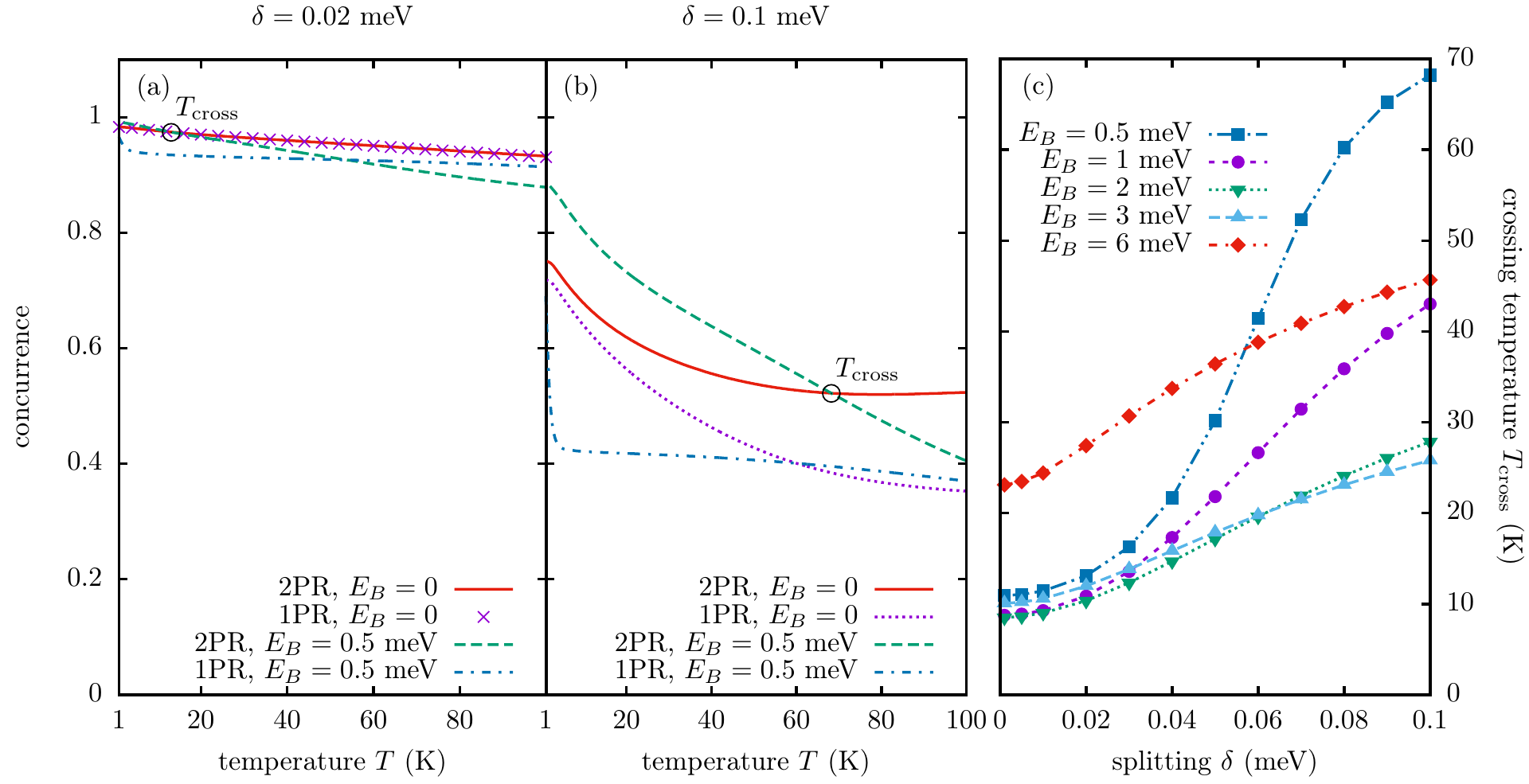}
\caption{Concurrence as a function of temperature for a fine-structure splitting $\delta=0.02$~meV [panel (a)] and $\delta=0.1$~meV [panel (b)] for all four
QD-cavity configurations introduced in Sec.~\ref{sec:configurations}.
Also shown is the crossing temperature $T_\text{cross}$ of the concurrence in the 2PR configuration with $E_B=0$ and the concurrence in the 2PR configuration 
with $E_B > 0$ as a function of $\delta$ for several values of the binding energy $E_B$ [panel (c)]. 
$T_\text{cross}$ is marked in panels (a) and (b) by a circle.}
\label{fig:all_config_temperature}
\end{figure*}

We conclude that the competition between single-photon and two-photon processes plays a decisive role for the concurrence. 
Furthermore, the arrangement of the cavity modes strongly affects the concurrence as one of the
competing processes can be either favored or suppressed. 
Finally, the values of the various detunings depend on the chosen configuration, resulting in
different effective phonon influences and
very different 
dependences of the concurrence on the exciton 
splitting for each of the considered QD-cavity configurations.

\subsection{Temperature dependence of the concurrence at a finite exciton splitting}
\label{subsec:temperature}
 
After the discussion in the last section it is clear that the temperature dependence of the concurrence
also differs for each of the four configurations. 
In this section we investigate in more detail the concurrence as a function of temperature for different fixed values of the exciton splitting.

Figure \ref{fig:all_config_temperature}(a) displays the concurrence as a function of the temperature for a typical value of the
exciton splitting $\delta=0.02$~meV, while a larger value $\delta=0.1$~meV is used in Fig.~\ref{fig:all_config_temperature}(b).  
As expected after the discussion of the 1PR configuration with a finite biexciton binding energy in Sec.~\ref{subsec:splitting}, 
the concurrence drops in this setting steeply for low temperatures followed
by a very weak $T$ dependence
compared with the other configurations for $T>4$ K and  both splittings.
Note that for $\delta>0$ in this configuration all sequential processes require the absorption
of phonons to bridge the energy mismatches and thus for $T\to0$ the phonon-free result should be reached.
Indeed, as seen  in Fig.~\ref{fig:all_config_temperature}(b), changing the temperature from 1 to 4 K
entails a very steep drop  of the concurrence before  it becomes almost independent of $T$ for $T>4$ K.

The 2PR configuration with a finite binding energy exhibits a rather strong temperature dependence for both selected values of $\delta$. 
Because of the weak influence of the exciton splitting in this configuration the concurrence reaches almost one for temperatures close to zero. 

The 2PR and 1PR configuration with a vanishing binding energy are, as discussed earlier, almost the same for small $\delta$. 
Therefore, for $\delta=0.02$~meV, the concurrence as a function of temperature is nearly identical for both configurations, with the 1PR 
result being marginally lower.
For the larger splitting $\delta=0.1$~meV these two configurations show a similar temperature dependence at very low temperatures but
at higher temperatures the concurrence decreases noticeably stronger in the 1PR configuration.

In the 2PR as well as in the 1PR configuration
the sequential single-photon processes are detuned on the order of $\delta$ when the biexciton binding energy is zero.
Since two-photon processes are more important in the 2PR configuration
the corresponding concurrence is higher for all temperatures than in the 1PR configuration when $E_{B}=0$.
However, this trend reverses for finite $E_B$ at high temperatures.

Let us now compare 2PR results with and without a finite biexciton binding energy.
As can be seen in Fig.~\ref{fig:all_config_temperature}(a) and Fig.~\ref{fig:all_config_temperature}(b), introducing a finite value for $E_{B}$ 
in the 2PR configuration leads to a higher concurrence only below a crossing temperature  which depends on $\delta$.
In fact, there is a crossing point of the 2PR concurrence evaluated at finite $E_{B}$ with each of the three other concurrences
considered here. It turns out that the setting with the lowest
crossing temperature is the 2PR configuration with vanishing biexciton binding energy.
We will denote the corresponding crossing temperature by $T_\text{cross}$ in the following.

For large splittings, a finite biexciton binding energy can raise the concurrence significantly
at low temperatures since the sequential single-photon 
emission processes become largely detuned and the importance of the two-photon processes is raised. 
Therefore, in the absence of phonons, a finite binding energy in general results in an increased concurrence,
a finding which was already proposed and 
discussed by Schumacher et al.\cite{Schumacher:2012}.

Above the crossing temperature, however, the pure dephasing coupling to the phonons alters this effect and a
finite binding energy in the 2PR configuration reduces the concurrence.
As the temperature increases, phonons raise the importance of the detuned single-photon processes because they become 
assisted by phonon absorption and emission. 
In the case of a finite binding energy, the cavity modes are more detuned from the electronic transitions involving exciton states. 
Therefore, the phonon spectral density $J(\omega)$ is probed at larger values so that the effective phonon coupling is
stronger compared with the situation 
without biexciton binding energy.
These two effects combined lead to a stronger decrease of the concurrence with increasing temperature for a finite $E_B$.
Thus, above $T_\text{cross}$, a finite biexciton binding energy reduces the concurrence and the protection of entanglement 
in the 2PR configuration is lost.

By comparing Fig.~\ref{fig:all_config_temperature}(a) and Fig.~\ref{fig:all_config_temperature}(b), one notices that the
crossing point of the 
concurrence in the 2PR configuration with and without a finite binding energy moves to lower temperatures
for smaller values of the exciton splitting.
For the concurrence, this means that the advantage provided by a finite $E_B$ is lost for small $\delta$ already
at low temperatures $T\sim 10$~K.
In Fig.~\ref{fig:all_config_temperature}(c) this crossing temperature is plotted against $\delta$
for several values of $E_B$.
For a given binding energy, $T_\text{cross}$ exhibits a monotonic increase with increasing exciton splitting
and converges to a finite value  of about 10~K in the limit $\delta \rightarrow 0$ and typical binding energies.
Therefore, for typical splittings on the order of several 10~\textmu eV, the protection of the entanglement due to
a finite binding energy is already lost at quite low temperatures.
As can be seen by comparing the results in Fig.~\ref{fig:all_config_delta}(a) and Fig.~\ref{fig:all_config_delta}(c),
the protection of the concurrence 
due to a finite $E_B$ in the 2PR configuration at temperatures close to zero improves for larger $\delta$. 
Thus, for larger $\delta$, higher temperatures are needed to destroy this
protection and $T_\text{cross}$ increases with increasing splitting for a given $E_B$.

Comparing the crossing temperature for different values of the binding energy, a nonmonotonic behavior is found at a given exciton splitting.
On the one hand, a higher value of the binding energy results in a better protection of the entanglement at temperatures close to zero.
But, on the other hand, the phonon influence and thus the influence of the temperature depends on the energy of the phonons needed to assist the detuned 
single-photon processes as the phonon spectral density $J(\omega)$ depends (nonmonotonically) on this energy.
In the case of the 2PR configuration with finite $E_B$, this roughly corresponds to half the binding energy.  
The nonmonotonic behavior of $T_\text{cross}$ as a function of $E_B$ at a given exciton splitting thus originates from the trade-off
between a better protection of the concurrence for higher binding energies at temperatures close to zero and the varying influence of the phonons due to the
nonmonotonic behavior of the phonon spectral density.
For example, the crossing point temperature in Fig.~\ref{fig:all_config_temperature}(c) for $E_B=1$~meV is always
higher than for $E_B=2$~meV. 
The reason is the much stronger temperature dependence in the latter situation as the phonon spectral density
is much higher for a phonon energy of $1$~meV 
than for a value of $0.5$~meV (cf.~Fig.~\ref{fig:spectral_density}).
However, at $\hbar\omega=1$~meV and $\hbar\omega=3$~meV the values of $J(\omega)$ are similar, which means that the phonon influence is similar for 
$E_B=2$~meV and $E_B=6$~meV and $T_{cross}$ is always higher in the latter case because of the
stronger protection due to the higher binding energy.

\section{Conclusion} 
\label{sec:conclusion}
We have analyzed how the competition between two-photon and single-photon emission processes as well as the coupling to LA phonons influences the degree of two-photon entanglement created in a QD-cavity system.
To this end we have calculated the concurrence of photon pairs simultaneously emitted in a biexciton-exciton cascade of a QD in a cavity for four different configurations.
We account for four electronic states (biexciton, two excitons, and the ground state), two degenerate orthogonally polarized cavity modes that are coupled to the electronic transitions, and cavity losses, as well as for a continuum of LA phonons coupled by the deformation potential interaction to the QD.
The numerical simulations are based on a path-integral scheme that allows the evaluation of quantities of interest without approximation to the model.

The four configurations considered in this paper comprise the two-photon resonant (2PR) and the one-photon resonant (1PR) configuration with a vanishing as well as a finite biexciton binding energy.
We find a wealth of interesting results and insights in the physics of the system at hand which we would like to briefly summarize below before we outline our main result at the end.

\begin{itemize}

\item[a)] The competition between two-photon and one-photon processes plays a decisive role for the concurrence and leads to strikingly different dependences on the exciton splitting $\delta$.

Among other things, we find, e.g., nonmonotonic dependences and deviations from the standard bell shape in the 2PR as well as in the 1PR configuration. 
While the 2PR and 1PR configuration without a biexciton binding energy lead to almost the same degree of entanglement for small splittings the 2PR configuration is favorable for larger splittings.

These results and the different dependences on the splitting $\delta$ can be very well explained by the different relative importance of direct two-photon and sequential single-photon contributions as well as the changing phonon impact when the resonance settings are varied.

\item[b)] The concurrence is in general only symmetric regarding the exciton splitting $\delta$ in the 2PR configurations. Additionally, LA phonons affect or even introduce the asymmetry in the 1PR configurations. Because of the characteristics of the phonon coupling this asymmetry is stronger at low temperatures as phonon absorption and emission processes are not equally likely to occur. 

\item[c)] The chosen configuration defines the detunings in the quantum dot-cavity system and results in different effective phonon influences and therefore also strongly different temperature dependences of the concurrence.

The 2PR and 1PR configuration with a vanishing binding energy have almost the same concurrence value and temperature dependence for the usual exciton splittings of several 10~\textmu eV.

The concurrence can be virtually independent of the temperature over a wide temperature range, as it is the case in the 1PR configuration with a finite binding energy and positive $\delta$ after the concurrence has fallen drastically with rising temperature for $T$ below 4 K.

\end{itemize}

In order to appreciate our main result, it should be noted that the 2PR configuration with finite biexciton binding energy has attracted a lot of attention \cite{Schumacher:2012,EdV,heinze2017} since this configuration has been proposed in order to reach high degrees of entanglement at finite fine-structure splittings.
The idea is that two-photon transitions are favored which are much less affected by the which-path information introduced by the fine-structure splitting than sequential single-photon processes.
Thus, a finite biexciton binding energy protects the entanglement from the destructive impact of the fine-structure splitting by making single-photon processes off-resonant.
Indeed, at low temperatures we find the highest degree of entanglement for this configuration which depends only little on the fine-structure splitting.
However, the concurrence in the 2PR configuration with finite biexciton binding energy exhibits a steep decrease with rising temperature, which can be explained by an enhanced interaction with phonons resulting from the frequency dependence of the phonon-spectral density combined with an increase of the importance of sequential single-photon processes at higher temperatures.

This strong temperature dependence is the origin of our most important result that for each of the other three considered configurations there is a finite temperature above which the corresponding concurrence is higher than in the 2PR case with finite biexciton binding energy.
Out of the configurations that we compare, the 2PR configuration with vanishing biexciton binding energy has the lowest such crossing temperature $T_\text{cross}$, which is found to depend on the fine-structure splitting as well as on the biexciton binding energy.
For splittings that are typically found in experiments on the order of several 10~\textmu eV or below and typical biexciton binding energies of few meV, $T_\text{cross}$ is around or even below 10~K.
Thus, the special distinction of the 2PR configuration with finite biexciton binding energy in terms of yielding the highest degree of entanglement for finite fine-structure splittings is lost already at rather low temperatures due to the phonon impact and the 2PR configuration with vanishing biexciton binding energy  becomes  more favorable for achieving the highest value of the concurrence.

\section*{Acknowledgments}

M. Cygorek thanks the Alexander-von-Humboldt foundation for support through a Feodor Lynen fellowship. 
A. Vagov acknowledges the support from the Russian Science Foundation under Project No. 18-12-00429, which was used to study dynamical processes non-local in time by the path-integral approach.
This work was also funded by the Deutsche Forschungsgemeinschaft (DFG, German Research Foundation) - project Nr. 419036043.

\bibliography{Bib_Paper2}
\end{document}